\documentclass[a4paper,fleqn,usenatbib]{mnras}

\usepackage[T1]{fontenc}
\usepackage{ae,aecompl}

\usepackage{graphicx}	
\usepackage{amsmath}	
\usepackage{amssymb}	

\title[VERITAS $\gamma$-Ray Observations of 1ES 2344+514]{Very-High-Energy $\gamma$-Ray Observations of the Blazar 1ES~2344+514 with VERITAS}

\author[S. O'Brien et al.]{
C.~Allen$^{1}$,
S.~Archambault$^{2}$,
A.~Archer$^{3}$,
W.~Benbow$^{4}$,
R.~Bird$^{5}$,\newauthor
E.~Bourbeau$^{2}$,
R.~Brose$^{6,7}$,
M.~Buchovecky$^{5}$,
J.~H.~Buckley$^{3}$,
V.~Bugaev$^{3}$,\newauthor
J.~V~Cardenzana$^{8}$,
M.~Cerruti$^{4}$,
X.~Chen$^{6,7}$,
J.~L.~Christiansen$^{1}$,
M.~P.~Connolly$^{9}$,\newauthor
W.~Cui$^{10,11}$,
M.~K.~Daniel$^{4}$,
J.~D.~Eisch$^{8}$,
A.~Falcone$^{12}$,
Q.~Feng$^{2}$,\newauthor
M.~Fernandez-Alonso$^{13}$,
J.~P.~Finley$^{10}$,
H.~Fleischhack$^{7}$,
A.~Flinders$^{14}$,
L.~Fortson$^{15}$,\newauthor
A.~Furniss$^{16}$,
G.~H.~Gillanders$^{9}$,
S.~Griffin$^{2}$,
J.~Grube$^{17}$,
M.~H\"utten$^{7}$,\newauthor
N.~H{\aa}kansson$^{6}$,
D.~Hanna$^{2}$,
O.~Hervet$^{18}$,
J.~Holder$^{19}$,
G.~Hughes$^{4}$,\newauthor
T.~B.~Humensky$^{20}$,
C.~A.~Johnson$^{18}$,
P.~Kaaret$^{21}$,
P.~Kar$^{14}$,
N.~Kelley-Hoskins$^{7}$,\newauthor
M.~Kertzman$^{22}$,
D.~Kieda$^{14}$,
M.~Krause$^{7}$,
F.~Krennrich$^{8}$,
S.~Kumar$^{19}$,\newauthor
M.~J.~Lang$^{9}$,
G.~Maier$^{7}$,
S.~McArthur$^{10}$,
A.~McCann$^{2}$,
K.~Meagher$^{23}$,\newauthor
P.~Moriarty$^{9}$,
R.~Mukherjee$^{24}$,
T.~Nguyen$^{23}$,
D.~Nieto$^{20}$,
S.~O'Brien$^{25}$\thanks{obrien.stephan@gmail.com},\newauthor
A.~O'Faol\'{a}in de Bhr\'{o}ithe$^{7}$,
R.~A.~Ong$^{5}$,
A.~N.~Otte$^{23}$,
N.~Park$^{26}$,
A.~Petrashyk$^{20}$,\newauthor
A.~Pichel$^{13}$,
M.~Pohl$^{6,7}$,
A.~Popkow$^{5}$,
E.~Pueschel$^{7}$,
J.~Quinn$^{25}$,\newauthor
K.~Ragan$^{2}$,
P.~T.~Reynolds$^{27}$,
G.~T.~Richards$^{23}$,
E.~Roache$^{4}$,
A.~C.~Rovero$^{13}$,\newauthor
C.~Rulten$^{15}$,
I.~Sadeh$^{7}$,
M.~Santander$^{24}$,
G.~H.~Sembroski$^{10}$,
K.~Shahinyan$^{15}$,\newauthor
I.~Telezhinsky$^{6,7}$,
J.~V.~Tucci$^{10}$,
J.~Tyler$^{2}$,
S.~P.~Wakely$^{26}$,
A.~Weinstein$^{8}$,\newauthor
A.~Wilhelm$^{6,7}$,
D.~A.~Williams$^{18}$
\\
$^{1}$Physics Department, California Polytechnic State University, San Luis Obispo, CA 94307, USA\\
$^{2}$Physics Department, McGill University, Montreal, QC H3A 2T8, Canada\\
$^{3}$Department of Physics, Washington University, St. Louis, MO 63130, USA\\
$^{4}$Fred Lawrence Whipple Observatory, Harvard-Smithsonian Center for Astrophysics, Amado, AZ 85645, USA\\
$^{5}$Department of Physics and Astronomy, University of California, Los Angeles, CA 90095, USA\\
$^{6}$Institute of Physics and Astronomy, University of Potsdam, 14476 Potsdam-Golm, Germany\\
$^{7}$DESY, Platanenallee 6, 15738 Zeuthen, Germany\\
$^{8}$Department of Physics and Astronomy, Iowa State University, Ames, IA 50011, USA\\
$^{9}$School of Physics, National University of Ireland Galway, University Road, Galway, Ireland\\
$^{10}$Department of Physics and Astronomy, Purdue University, West Lafayette, IN 47907, USA\\
$^{11}$Department of Physics and Center for Astrophysics, Tsinghua University, Beijing 100084, China.\\
$^{12}$Department of Astronomy and Astrophysics, 525 Davey Lab, Pennsylvania State University, University Park, PA 16802, USA\\
$^{13}$Instituto de Astronomia y Fisica del Espacio, Casilla de Correo 67 - Sucursal 28, \\~~~~~(C1428ZAA) Ciudad Autónoma de Buenos Aires, Argentina\\
$^{14}$Department of Physics and Astronomy, University of Utah, Salt Lake City, UT 84112, USA\\
$^{15}$School of Physics and Astronomy, University of Minnesota, Minneapolis, MN 55455, USA\\
$^{16}$Department of Physics, California State University - East Bay, Hayward, CA 94542, USA\\
$^{17}$Department of Physics, Stevens Institute of Technology, Hoboken, NJ 07030, USA\\
$^{18}$Santa Cruz Institute for Particle Physics and Department of Physics, University of California, Santa Cruz, CA 95064, USA\\
$^{19}$Department of Physics and Astronomy and the Bartol Research Institute, University of Delaware, Newark, DE 19716, USA\\
$^{20}$Physics Department, Columbia University, New York, NY 10027, USA\\
$^{21}$Department of Physics and Astronomy, University of Iowa, Van Allen Hall, Iowa City, IA 52242, USA\\
$^{22}$Department of Physics and Astronomy, DePauw University, Greencastle, IN 46135-0037, USA\\
$^{23}$School of Physics and Center for Relativistic Astrophysics, Georgia Institute of Technology, \\~~~~~837 State Street NW, Atlanta, GA 30332-0430\\
$^{24}$Department of Physics and Astronomy, Barnard College, Columbia University, NY 10027, USA\\
$^{25}$School of Physics, University College Dublin, Belfield, Dublin 4, Ireland\\
$^{26}$Enrico Fermi Institute, University of Chicago, Chicago, IL 60637, USA\\
$^{27}$Department of Physical Sciences, Cork Institute of Technology, Bishopstown, Cork, Ireland
}
\date{Accepted 2017 July 10. Received 2017 June 24; in original form 2016 Novemeber4}

\pubyear{2017}

\begin{document}
\label{471}
\volume{471}
\pagerange{2117--2123}
\maketitle

\begin{abstract}
We present very-high-energy $\gamma$-ray observations of the BL Lac object 1ES~2344+514 taken by the Very Energetic Radiation Imaging Telescope Array System (VERITAS) between 2007 and 2015. 
	1ES~2344+514 is detected with a statistical significance above background of $20.8\sigma$ in $47.2$ hours (livetime) of observations, making this the most comprehensive very-high-energy study of 1ES~2344+514 to date.
	Using these observations the temporal properties of 1ES~2344+514 are studied on short and long times scales. 
	We fit a constant flux model to nightly- and seasonally-binned light curves and apply a fractional variability test, to determine the stability of the source on different timescales. 
	We reject the constant-flux model for the 2007-2008 and 2014-2015 nightly-binned light curves and for the long-term seasonally-binned light curve at the $> 3\sigma$ level. 
	The spectra of the time-averaged emission before and after correction for attenuation by the extragalactic background light are obtained. 
	The observed time-averaged spectrum above 200 GeV is satisfactorily fitted {(${\chi^2/NDF = 7.89/6}$)} by a power-law function with index $\Gamma = 2.46 \pm 0.06_{stat} \pm 0.20_{sys} $ and extends to at least 8~TeV. 
	The extragalactic-background-light-deabsorbed spectrum is adequately fit (${\chi^2/NDF = 6.73/6}$) by a power-law function with index $\Gamma = 2.15 \pm 0.06_{stat} \pm 0.20_{sys} $ while an F-test indicates that the power-law with exponential cutoff function provides a marginally-better fit ($\chi^2/NDF $ = $2.56 / 5 $) at the 2.1$\sigma$ level. 
	The source location is found to be consistent with the published radio location and its spatial extent is consistent with a point source. 
\end{abstract}



\begin{keywords}
astroparticle physics -- gamma-rays: galaxies -- BL Lacertae objects: 1ES~2344+514 = VER~J2347+517 
\end{keywords}



\section{Introduction}

1ES~2344+514 \citep[R.A.: 23h~47$\arcmin$~04.837$\arcsec$, Dec.:  +51$^\circ$~42$\arcmin$~17.881$\arcsec$ (J2000),][]{2008AJ....136..580P} is a BL Lac object located at a redshift of $z=0.044$ \citep{Perlman96}. It was the third reported
extragalactic source of very-high-energy (VHE, E$\gtrsim100$~GeV) $\gamma$-rays \citep{Catanese98} after Markarian 421 and Markarian 501. To date, approximately 60 BL Lac objects have been detected as VHE sources.\footnote{\url{http://tevcat.uchicago.edu/}} BL Lac objects, along with flat spectrum radio quasars, constitute the blazar class of active galactic nuclei (AGN), which are thought to consist of AGN with relativistic jets that happen by chance to be oriented towards the Earth.

 BL Lacs are highly variable and exhibit double-humped broadband non-thermal spectra, with the first peak, located in the infra-red to X-ray regime, believed to be relativistically beamed incoherent synchrotron radiation. The location of the synchrotron peak is used to classify a BL Lac as a low- (LBL), intermediate- (IBL) or high-frequency-peaked BL Lac object (HBL) \citep{Padovani1995ApJ}. 1ES~2344+514 is classified in the HBL category, as are
the majority ($\sim$80\%) of the VHE-detected BL Lacs to date. 
The higher-energy peak, located in the $\gamma$-ray regime, is likely due to inverse-Compton scattering of either the synchrotron photons themselves (known as \emph{synchrotron self-Compton}, SSC) or soft photons external to the jet (known as \emph{external inverse-Compton}, EIC),
or a combination of both, although other emission mechanisms such as from hadronic processes have not been ruled out \citep[for an overview see, e.g.][]{bottcher2007modelling}. Measurements of the $\gamma$-ray variability and spectrum of a BL Lac object, when combined with multi-wavelength measurements, may be used to constrain the size of the emission region, and to test theoretical models and constrain model parameters such as the magnetic field strength and electron energy distribution \citep[e.g.][]{Cerruti13}. It has been found that most HBLs are adequately described by pure single zone SSC models while IBLs and LBLs require increasing EIC contributions to reproduce the observed spectra (see \citet{bottcher2007modelling} and references therein). Spectral studies of VHE BL Lac objects tend to utilize bright states associated with flares due to the high photon statistics obtained. Long-term monitoring, as presented here for 1ES 2344+514, is needed to obtain statistically-significant detections over a reasonable energy range to facilitate the study and spectral modeling of low-flux states. In addition, long-term monitoring is important for characterizing the duty cycle of flaring blazars, which has an impact on the contributions of blazars to the extragalactic $\gamma$-ray background radiation fields \citep{giommi2006backgrounds,pittori2007dutycycle}.

VHE photon fluxes undergo energy-dependent attenuation when travelling cosmological distances due to pair-production via interaction with low-energy optical/IR extragalactic background light (EBL) photons. Thus, the observed VHE spectrum of an object is the emitted spectrum modified by absorption due to EBL photons. EBL absorption ultimately limits the distance over which VHE $\gamma$-ray sources can be detected. Correcting for EBL absorption is necessary in order to uncover the intrinsic VHE spectrum of a BL Lac object. The intensity and spectrum of the EBL is also of cosmological interest as it contains a record of the star-formation history of the Universe, but direct measurement is difficult due to foreground contamination by zodiacal light. However, constraints can be obtained from VHE spectral measurements of distant objects by finding the maximum allowed EBL such that the de-absorbed spectrum does not conflict with the hardest allowed spectrum from theoretical arguments or extrapolated spectra from lower energies where EBL absorption does not take place. \citep[For a review see, e.g.][]{Dwek13}. The most comprehensive studies involve combining the constraints from multiple different objects at a variety of redshifts \citep{DAW_EBL}. In this work we use a published EBL model to estimate the intrinsic VHE spectrum of 1ES~2344+514.

If BL Lac objects generate multi-TeV emission then the interactions with EBL photons may produce magnetically-broadened cascades resulting in extended halo emission around these sources. Spatial profiles of such objects in VHE $\gamma$-rays may be used to place model-dependent constraints on intergalactic magnetic fields \citep[e.g.][]{Aharonian1994halos, archambault2017igmf}. While 1ES~2344+514 is not  an ideal candidate for magnetically-broadened emission due to its proximity and relatively low flux (an estimate of the cascade fraction is below the VERITAS sensitivity), a search for spatially-extended emission was carried out as part of this study to compare the location of the site of the VHE emission with that at other wavelengths.

1ES~2344+514 is an important target for VERITAS. The source is regularly monitored as part of the blazar science program \citep{Benbow_ICRC15} in order to obtain measurements of its long-term flux states. 
These observations can be used to build statistics for spectral determination at the highest energies, and to trigger intensive multiwavelength observations should exceptional flaring activity be observed. 
The spectral information obtained from these observations will be used as part of the VERITAS multi-object study to place constraints on the EBL.
In this publication we report on the analysis of the complete VERITAS data set taken on 1ES~2344+514 from October 2008 to January 2015. The data from 2007-2008, previously published in \citet{Acciari11}, are also incorporated in order to provide updated spectra and light curves. 
These data provide important measurements of the highest energy component emission from 1ES~2344+514, which can be combined with multiwavelength observations to search for correlated variability and to construct the overall SED for model evaluation.

\section{1ES~2344+514 VHE $\gamma$-ray Observational History}

1ES~2344+514 was first detected as a VHE $\gamma$-ray source by the Whipple 10m telescope during observations made between 1995 and 1996 \citep{Catanese98}, with most of the signal coming from a flare on a single night. These and subsequent observations performed by several imaging atmospheric Cherenkov telescopes (IACTs) \citep[e.g.][]{Aharonian04,Godambe07,Albert07,Acciari11, Aleksic13} have revealed the source to have variable flux states. The integral flux is typically less than 10\% that of the Crab Nebula, but day-scale flares of 60\% and 50\% of the Crab Nebula flux have been observed
\citep{Catanese98,Acciari11}.

Current-generation IACTs have allowed variability and spectral studies of 1ES~2344+514 to be extended to lower flux states.  The MAGIC collaboration observed 1ES~2344+514 in a particularly low state for $\sim$20 hours spread across 14 nights (2008 October 20 to 2008 November 30) as part of a multiwavelength campaign; it was marginally detected at a level of 3.5 standard deviations above the background ($\sigma$) and  determined to have a flux approximately 2.5\% that of the Crab Nebula with a power-law energy spectrum of index $2.4 \pm 0.4_{stat} $ in the energy range from 0.17 TeV to 2 TeV \citep{Aleksic13}. VERITAS detected 1ES 2344+514 at the 20$\sigma$ level in 18 hours of observations spread across 37 nights in 2007/2008 (from October 4, 2007 to January 11, 2008) including a one-day flare of $\sim$50\% of the Crab Nebula flux on December 7, 2007 (MJD 54441)
\citep{Acciari11}. Analysis of the data excluding the flare revealed a low-state flux of $\sim$7.6\% of the Crab Nebula with a spectrum well fitted by a power law of index $2.78 \pm 0.09_{stat} \pm 0.15_{sys}$ in the energy range 0.39 to 8.3 TeV. The flare-state spectrum was well fitted by a power law of index $2.43 \pm 0.22_{stat} \pm 0.20_{sys}$, consistent with the spectral analysis of the previous large flare seen with the Whipple 10m telescope on 1995 December 20 which showed a power-law index of $2.54 \pm 0.17_{stat} \pm 0.07_{sys}$ \citep{Schroedter05}.  MAGIC observations in 2005/2006 (27 nights in the interval 2005 August 3 to 2006 January 1) totalling 32 hours provided an 11$\sigma$ detection with an average flux of $\sim$10\% that of the Crab Nebula, with a spectrum well fitted by a power law of index $2.95 \pm 0.12_{stat} \pm 0.2_{sys}$ between 0.14 and 5 TeV \citep{Albert07}. HBL objects tend to show a harder spectral index with increased flux, with the MAGIC observations in 2008 being an unusual exception as discussed by \cite{Aleksic13}.

\section{VERITAS Observations and Analysis}

The Very Energetic Radiation Imaging Telescope Array System \citep[VERITAS,][]{Holder06} is an array of four IACT telescopes, located at the Fred Lawrence Whipple Observatory in southern Arizona. Each telescope is of 12m diameter, with a 499-pixel photomultiplier-tube (PMT) camera located at the focal plane. Since commissioning in 2007 the array has undergone two major upgrades to improve performance: in May 2009 Telescope 1 was 
moved to improve the collection area of the array, which, along with upgraded mirror alignment \citep{mccann2010new} and trigger systems \citep{zitzer8360veritas}, resulted in a 30\% improvement in sensitivity  \citep{Perkins09}. In summer 2012 the cameras were upgraded with higher-quantum-efficiency PMTs which resulted in significantly-improved sensitivity below 100~GeV  \citep{Otte11}. In its current configuration VERITAS can detect a source with a flux of 1\% that of the Crab Nebula at the 5$\sigma$ level in $\sim$25~hours, has a minimum individual photon angular resolution (68\% containment radius) of $\sim 0.1^\circ$ at 1 TeV, an energy resolution of 15-25\% and an effective area for $>$1~TeV photons on the order of $10^5$~m$^2$. For a detailed discussion of the current performance of VERITAS see \cite{Park15}.

In addition to normal dark-sky observations, VERITAS has been operated under moderate moonlight using reduced high voltage (RHV) to the PMTs, resulting in
increased observing time, but with a loss of sensitivity below $\sim$200~GeV \citep{Griffin15}. The extra observing time that RHV observations offer has already been beneficial with the detection of a flare from the blazar 1ES~1727+502 \citep{Archambault15}. 

The 47.2 hours of data presented here include all three-telescope configurations, incorporate 11.6 hours of RHV observations and include the data presented in \cite{Acciari11}. The flare (on MJD 54441) reported in \cite{Acciari11} was observed under less-than-optimal weather conditions and is excluded from this analysis, but is shown in the light curves in order to put the long-term variability of the source in perspective.

Independent analyses were carried out using the collaboration's two main data analysis software packages and excellent agreement was found. Details of the analysis procedures used by VERITAS can be found in \cite{Acciari08_LSI}.
The $\gamma$-ray selection criteria used were optimised for a source with 5\% Crab Nebula flux and a power-law spectral index of 2.5.  Background estimation was performed using the reflected-region method \citep{Fomin94} with eight reflected regions, and the significance of the $\gamma$-ray excess was calculated using Equation 17 of \citet{LiandMa83}. 

We define the nominal VERITAS observing season to be from September to June, with  September to August being a full year. The breakdown of the observations by observing season is given in Table \ref{tab_Flux}. To compare the measured fluxes to that of the Crab Nebula we use the differential spectrum derived in \cite{Hillas98}. Due to the varying energy thresholds of the observations caused by zenith angle differences, the inclusion of RHV data, and different array configurations, we have quoted the detection significance and conducted the light-curve analysis above a common energy threshold of 0.35 TeV. It should be noted that this energy threshold cut is not applied to the spectral analysis.

%
\begin{table}
\renewcommand*{\thefootnote}{\alph{footnote}}
\renewcommand{\arraystretch}{1.3}
\caption{Breakdown of observations by season including hours of data (livetime), detection significance, flux, the results of a $\chi^2$ test for constant emission, and the results of the fractional variability ($F_{var}$) test (see Section \ref{section:variability}).  
}
\resizebox{\linewidth}{!}{
\begin{tabular}{ p{1.5cm} | p{1.cm}  | p{1.cm} |  c | c | c }
Data Set  &  Livetime  & Detection& Flux\textsuperscript{$\dagger$}   & $\chi^2$/NDF  & $F_{var}$\\ 
         &  (hours) & ($\sigma$) & ($>0.35$~TeV)  &  & \\ 
\hline

2007-2008 & $	17.1	$ & $16.2	$  & $13.2 \pm 1.0 $ &	$192 / 37$  & $0.78 \pm 0.13$ \\
2008-2009 & $	0.4 	$ & $-0.7	$ &  $<6.4 \text{}  $ &	  N/A & N/A\\
2010-2011 & $	5.4 	$ & $3.3	$  & $3.1	\pm 1.0  $ &	$26 / 15$  & $0.56 \pm 0.66$\\
2011-2012 & $	3.0 	$ & $2.6	$ &  $2.8	\pm 1.2  $ &	$23 / 9 $  & $1.58 \pm 0.50$\\
2012-2013 & $	4.2 	$ & $9.7	$ &  $8.3	\pm 1.2  $ &	$19 / 8 $  & -\textsuperscript{$\ddagger$}\\
2013-2014 & $	7.8 	$ & $2.5	$ &  $1.5	\pm 0.6  $ &	$16 / 12$  & -\textsuperscript{$\ddagger$}\\
2014-2015 & $	9.3	$ & $11.0	$ &  $6.4	\pm 0.7  $ &	$52 / 10$  &  $0.52 \pm 0.23$\\\hline
2007-2015 & $	47.2	$ & $20.8  $ &  $7.1  \pm 0.4  $ &  $464 / 97$  & $0.97 \pm 0.10$     \\\hline
\multicolumn{6}{l}{\textsuperscript{$\dagger$}\footnotesize{units:  $\times10^{-12}\ \text{cm}^{-2}\ \text{s}^{-1}$}} \\ 
\multicolumn{6}{l}{\textsuperscript{$\ddagger$}\footnotesize{undefined as variance determined to be smaller than the mean error - see section \ref{section:variability}. }}
\end{tabular}
}
\label{tab_Flux}

\end{table}


\section{Results}

\subsection{Overall Detection}
\par The livetime of 47.2 hours on 1ES 2344+514 results in a total number of `on' and `off' region counts of 1232 and 4826 respectively (using eight background regions and a common energy threshold of 0.35 TeV), hence we report a detection significance of 20.8$\sigma$ above background.

 The time-averaged event rate for all nights is measured to be $(0.20 \pm 0.01)\ \gamma\,\text{min}^{-1}$  and the total 
 time-averaged flux above $0.35$~TeV is calculated to be $F(E \text{\textgreater} 0.35\ \text{TeV}) = (7.1 \pm 0.4)\times10^{-12} \ \text{cm}^{-2}\ \text{s}^{-1}$ or $7 \%$ that of the Crab Nebula.

\subsection{Source Localisation}
\par  A symmetric two-dimensional Gaussian function is fitted to the excess-counts sky map using a 
$\chi^2$ method. The $\chi^2$ statistic is minimised at the coordinates (J2000):\\ 
\hspace*{4em} R.A.: (23h 47$\arcmin$ 4$\arcsec$) $\pm$ (0h 0$\arcmin$ 2$\arcsec$ $)_{stat}$,\\ 
\hspace*{4em} Dec.:  (+51$^\circ$ 42$\arcmin$ 49$\arcsec$) $\pm$ (0$^\circ$ 0$\arcmin$ 16$\arcsec$ $)_{stat}$\\ 
The fitted extension is consistent with a point source convolved with the point-spread function of VERITAS. The systematic error on the VERITAS pointing accuracy is $< 25 \arcsec$, which gives systematic errors on the right ascension of  
$\pm$(0h~0$\arcmin$~3$\arcsec$) and on the declination of $\pm$(0$^\circ$~0$\arcmin$~25$\arcsec$). As this is the first study of the source location with VERITAS, {1ES~2344+514} is allocated the VERITAS catalogue name VER~J2347+517.

\subsection{Variability Analysis}
\label{section:variability}
\par
\begin{figure*}
   \centering
   \includegraphics[width=14cm]{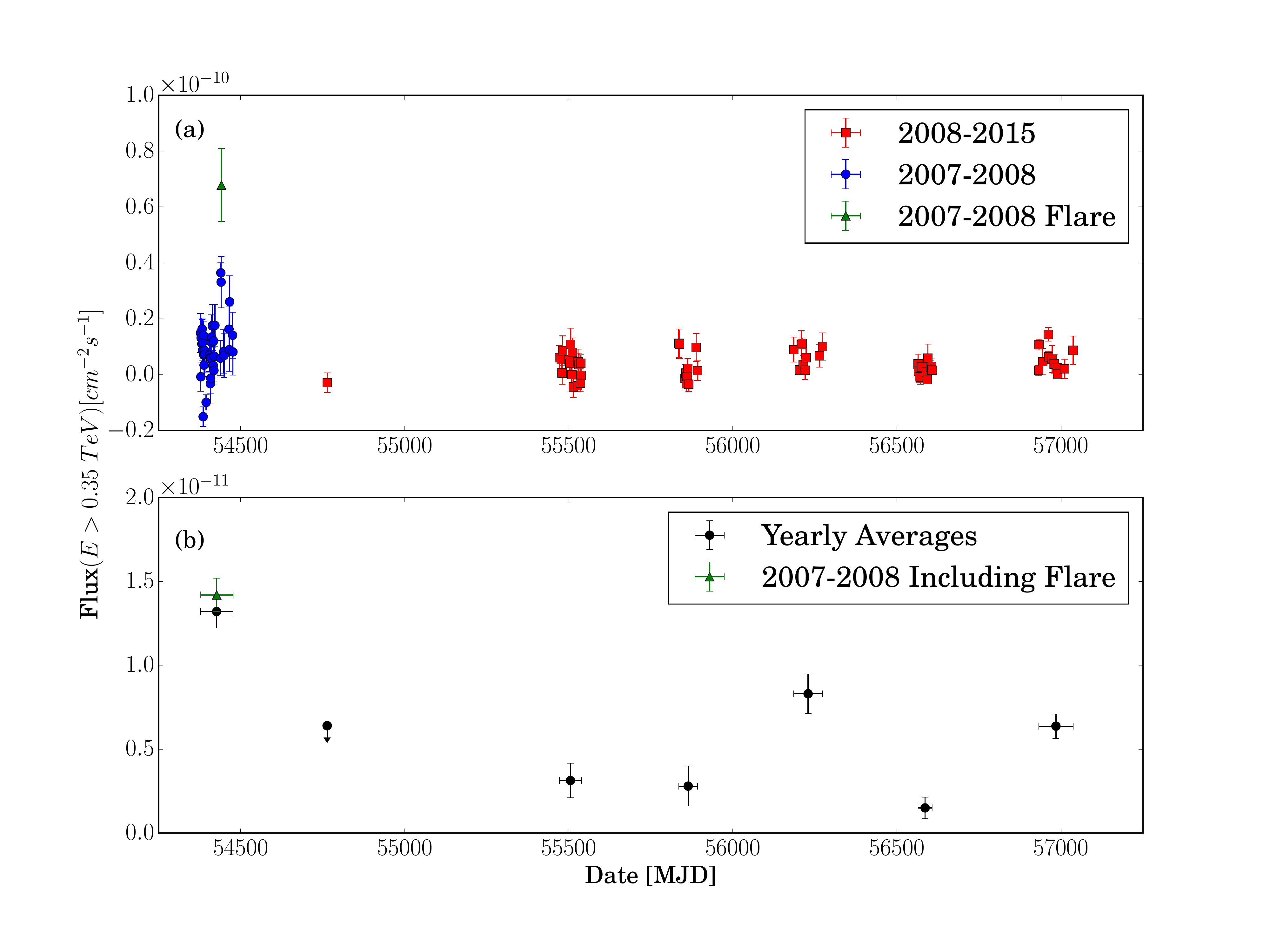}
      \caption{Light curves for {1ES~2344+514}. Panel (a) shows the average nightly flux. Observations taken during the 2008-2015 seasons are plotted as red squares, observations taken during the 2007-2008 season are plotted as blue circles. All points are plotted regardless of their significance. Panel (b) shows the average flux within each seasonal bin with an upper limit plotted for the 2008-2009 observation season. In calculating average fluxes, we consider all flux measurements regardless of their significance to obtain an unbiased measurement.}  
    \label{fig_LightCurve}
\end{figure*}

 We tested the stability of the season-to-season flux by applying a $\chi^2$ test for constant emission to the seasonally-averaged fluxes. We also tested for variability within each season by binning 
the data by night and performing a $\chi^2$ test for constant emission. The results of the $\chi^2$ tests are summarised in Column 5 of Table \ref{tab_Flux} and the light curves are shown in Figure \ref{fig_LightCurve}. The flare on the night of MJD 54441 is excluded from this analysis, but included in
Figure \ref{fig_LightCurve} for comparison reasons.
Due to having only $\sim$30 minutes of observations during the 2008-2009 season we exclude that season from the seasonal variability test and calculate a 95\% upper limit on the flux. 
We reject the constant-flux hypothesis on seasonal timescales due to the calculated $\chi^2/NDF$ of $121/5$. Within each season, the light curves were found to be inconsistent with the constant-flux model at a level >3$\sigma$ for the 2007-2008 and  2014-2015 observing seasons.

\par The fractional variability statistic \citep[$F_{var}$,][]{vaughan2003characterizing} given by
$$ F_{var}= \sqrt{ \frac { S^2 - \overline{\sigma_{err}^2}} { \overline{x}^2 } }\pm 
\sqrt{ \left( \sqrt{\frac{1}{2N}} \frac{ \overline{\sigma_{err}^2} }{\overline{x}^2 F_{var}}\right)^2 + \left( \sqrt{ \frac{\overline{\sigma_{err}^2}}{N} }\frac{1}{\overline{x}}\right)^2 } , $$
 where $S$ is the sample variance, $\overline{\sigma_{err}^2}$ is the  mean squared error on the flux of the sample, $N$ is the sample size and  $\overline{x}$ is the average flux of the sample, is calculated for each season.  The calculated $F_{var}$ indicates the fractional observed variance in excess of the known average statistical uncertainty.  The results are summarised in Column 6 of Table \ref{tab_Flux}. Variability in excess of the statistical uncertainty is observed in the 2007-2008, 2011-2012 and 2014-2015 observing seasons, while the 2010-2011, 2012-2013 and 2013-2014 seasons exhibit variability which is consistent with that expected from statistical uncertainties.

\par For consistency, both the already-published 2007-2008 and newly-presented 2008-2015 data sets are analysed using the most recent version of the VERITAS analysis software. This allows for a comparison of the two data sets.
The time-averaged flux for the low-state observations during the 2007-2008 season is measured to be 
$ F(E \text{\textgreater} 0.35\ \text{TeV}) = (13.2 \pm 1.0)\times10^{-12}\ \text{cm}^{-2}\ \text{s}^{-1}$, 
equivalent to $13 \%$ that of the Crab Nebula, while the 2008-2015 data set shows a considerably dimmer flux level of  $F(E \text{\textgreater} 0.35\ \text{TeV}) = (4.4 \pm 0.4 )\times10^{-12}\ \text{cm}^{-2}\ \text{s}^{-1}$, 
equivalent to $4\%$ that of the Crab Nebula.  The significance of the decrease is calculated as 
$(F_{2007-2008} - F_{2008-2015})/\sqrt{\delta F_{2007-2008}^2 + \delta F_{2008-2015}^2}$, where $F$ and $\delta F$ are the flux and flux error for the respective observation periods and has the value $8.3 ~\sigma$.

\subsection{Spectral Analysis}
\par To investigate any possible bias introduced by combining data with different energy thresholds, we compare the best fit energy spectra before and after applying the common energy threshold of 350~GeV. The resulting best-fit spectra are found to be in excellent agreement, indicating no bias is present. The spectral analysis was thus conducted above the lower energy threshold of 200~GeV.  

\par The observed and EBL-corrected differential spectra were fitted with power-law functions, and, in addition, the
EBL-corrected spectrum was fitted with a power-law with exponential cut-off function to test the intrinsic spectral curvature.

The  power-law  function fitted is of the form 
$$ \frac{dN}{dE} = N_0 \left(\frac{E}{E_0}\right)^{-\Gamma} ,$$ 
where $E_0$ is the decorrelation energy (i.e.\ the energy at which the correlation between the normalisation flux ($N_0$) and the spectral index ($\Gamma$) is a minimum). The fit is applied to the original low-state data set published in \citet{Acciari11}, the new data
taken since that publication, and the combined data set, with the results presented in Table \ref{tab_Spectrum}. 

The systematic errors quoted are the standard VERITAS errors of 20\% on the flux normalisation and 0.20 on the index for a power-law function  \citep{Madhavan2013PhD}. The 2007-2008 dataset is found to be in good agreement with the previously published spectrum, when statistical and systematic uncertainties are considered.

\begin{figure}
\centering
\includegraphics[width=9cm]{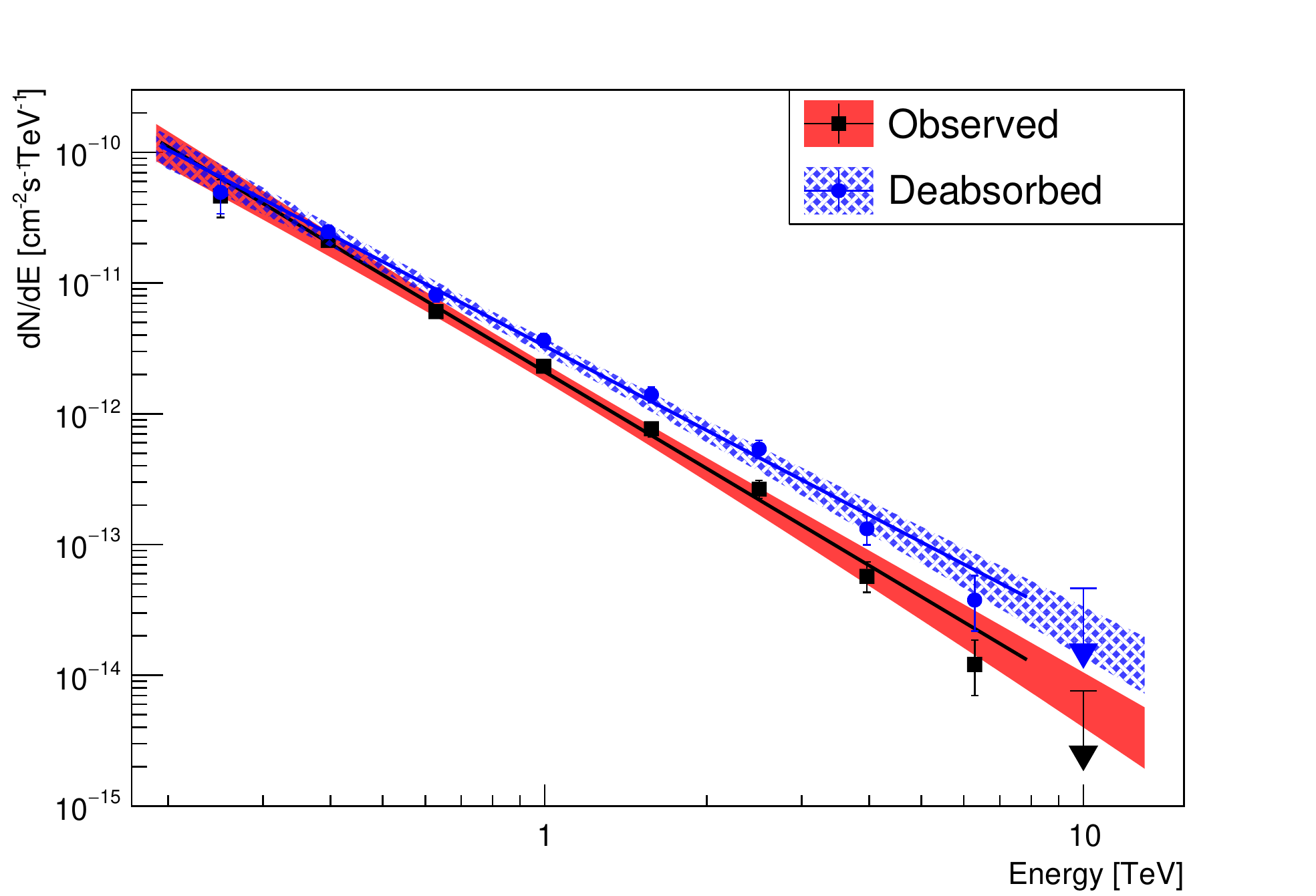}
\caption{The differential energy spectrum of {1ES~2344+514} fit with a power law with a 95\% statistical confidence band plotted as a shaded region. The deabsorbed energy spectrum is obtained using the \protect\cite{franceschini2008extragalactic} EBL model. Details of the fits can be found in Table \ref{tab_Spectrum}.  Points failing to meet a threshold significance of 2 $\sigma$ or  with less than  5 `on' events are plotted as upper limits and are not included in the fit. }       
\label{fig_Spectrum}

\end{figure}

The spectral points for the combined 2007-2015 data set, excluding the flare night of MJD 54441, along with the best-fit power law and the 95\% confidence interval (statistical errors only)
on the fit, are shown in Figure \ref{fig_Spectrum}. Points are shown when the significance exceeds
2$\sigma$ in a bin and there are at least five `on' events, otherwise uppers limit are presented.
We extrapolate the confidence interval obtained from the best-fit function out to the final energy bin (between $8\ \text{TeV}$ and $13\ \text{TeV}$) and find agreement between the model and upper limit.

To investigate the intrinsic spectrum of the source we deabsorbed the observed 2007-2015 spectrum using the EBL model of \cite{franceschini2008extragalactic}. This deabsorbed spectrum is then fitted with a power-law function (see Table \ref{tab_Spectrum}) and a power-law with exponential cut-off function of the form:
$$ \frac{dN}{dE} = N_0 \left(\frac{E}{E_0}\right)^{-\Gamma} e^{-\frac{E}{E_{c}} } , $$ 
where $E_{c}$ is the cut-off energy. The power law with exponential cut-off is well fitted with a $\chi^2/NDF $ = $2.56 / 5 $ and parameters $N_0 = (5.19 \pm 1.99_{stat}) \times 10^{-13} \text{cm}^{-2}\,\text{s}^{-1}$ , $\Gamma = 1.82 \pm 0.19_{stat} $, $E_0 = 3.3 \text{ TeV} $  and $E_{c} = 4.38 \pm 2.42_{stat} \text{ TeV} $. To test whether the power-law with exponential cut-off model provides a significantly-improved fit we apply a nested F-test, which results in an F-statistic of 8.13, corresponding to a 2.1$\sigma$ improvement over the power-law model.

\begin{table}
\renewcommand*{\thefootnote}{\alph{footnote}}
\renewcommand{\arraystretch}{0.5}
\centering
\caption{Spectral analysis of {1ES~2344+514} broken-down into observation periods. The systematic errors on $N_0$ and $\Gamma$ are estimated to be $20\%$ and $0.20$ respectively.}
\resizebox{\columnwidth}{!}{
\resizebox{12cm}{!}{
\begin{tabular}{ c | c  | c |  c | c  }

          \multicolumn{5}{c}{Power-law fit}  \\\hline
Data Set & $N_0$ & $E_0$ & $\Gamma$ & $\chi^2$/NDF \\
         & ($\times10^{-12}\ \text{cm}^{-2}\,\text{s}^{-1}$) & (TeV)  &    \\ \hline
&&&&\\ 2007-2008 & $ 3.19 \pm 0.23_{stat} $ & 1.11 & $2.43 \pm 0.09_{stat} $ & 4.03/5  \\ 
&&&&\\ 2008-2015 & $ 3.42 \pm 0.29_{stat} $ & 0.69 & $2.63 \pm 0.14_{stat} $ & 3.39/4\\ 
&&&&\\ \hline
&&&&\\ 2007-2015 & $ 2.65 \pm 0.14_{stat} $ & 0.91 & $2.46 \pm 0.06_{stat} $ & 7.89/6 \\
&&&&\\  2007-2015\textsuperscript{$\dagger$} 
                 & $ 4.04 \pm 0.22_{stat} $ & 0.91 & $2.15 \pm 0.06_{stat}$ & 6.73/6 \\&&&&\\ \hline
\multicolumn{5}{l}{\raisebox{-1.5ex}{\textsuperscript{$\dagger$}\footnotesize{ Deabsorbed using the \protect\cite{franceschini2008extragalactic} EBL model.}}} \\
\end{tabular}
 } 
 }

\label{tab_Spectrum}
\end{table}

\section{Discussion and Conclusions}

\par We have presented the results of VERITAS observations of 1ES~2344+514 using all good-weather data
taken between 2007 and 2015, including data taken under moderate moonlight with a reduced high-voltage supplied to the PMTs.   This extended data set incorporates 30.1 additional hours of monitoring data beyond the 2007 data set already published by VERITAS in \citet{Acciari11}. 1ES~2344+514 is detected at greater than 20$\sigma$ above background
in the overall data set, with the mean 2008-2015 flux showing a significant ($8.3\sigma$) $\sim$70\% decrease from $\sim$13\% to $\sim$4\% of the Crab Nebula flux. There is evidence for flux variability from season to season, and within seasons, but no new significant flaring activity was observed. The observations, when combined with other VHE observations, indicate that there is not a steady baseline flux level for the VHE emission from this object. The observed \emph{flickering} variability is not uncommon for BL Lac objects and indicates that the VHE emission is not from a large scale structure within the galaxy, but instead from a compact region most probably within the jet, with variability possibly caused by turbulence in the jet or variations in the electron supply \citep[see, for example,][]{Howard2004AJ}.


The first study of the location and extension of the VHE $\gamma$-ray emission with VERITAS of 1ES~2344+514 was conducted. The best-fit coordinates  (J2000) were found to be:\\
\hspace*{1em} R.A.: (23h 47$\arcmin$ 4$\arcsec$) $\pm$ (0h 0$\arcmin$ 2$\arcsec$ $)_{stat}$ $\pm$ (0h 0$\arcmin$ 3$\arcsec$ $)_{sys}$,\\ 
\hspace*{1em} Dec.:  (+51$^\circ$ 42$\arcmin$ 49$\arcsec$) $\pm$ (0$^\circ$ 0$\arcmin$ 16$\arcsec$ $)_{stat}$ $\pm$ (0$^\circ$ 0$\arcmin$ 25$\arcsec$ $)_{sys}$.\\ 
 The location of the source is consistent with the radio location of R.A.: 23h~47$\arcmin$~04.837$\arcsec$, Dec.:  +51$^\circ$~42$\arcmin$~17.881$\arcsec$ (J2000), reported by \cite{2008AJ....136..580P} when the statistical and systematic errors are taken into account. The fitted extension is consistent with a point source convolved with the
 point-spread function of VERITAS, revealing no evidence for extended halo-type emission that might be expected 
 from magnetically broadened cascades. A recent dedicated VERITAS analysis of the spatial extents of other BL Lac objects also found no evidence for extended emission \citep{archambault2017igmf}, so this result on 1ES~2344+514 is not unexpected.

\par  The observed time-averaged spectrum is best fitted by a power law with index of $\Gamma = 2.46 \pm 0.06_{stat} \pm 0.20_{sys} $.  The spectrum extends to at least 8 TeV, and is approaching a region where absorption due to the EBL is starting to become significant (the optical depth for 8 TeV photons due to EBL absorption is 1.3 for a source at a redshift of z=0.044 according to the \citet{franceschini2008extragalactic} model). The EBL-deabsorbed spectrum was fitted with both a power law and a power law with exponential cutoff; the power law gave a satisfactory fit ($\chi^2/NDF$ of 6.73/6) with 
a spectral index of $\Gamma = 2.15 \pm 0.06_{stat} \pm 0.20_{sys}$, while the power law with exponential cutoff 
yielded only a marginal (2.1$\sigma$) improvement to the fit. 
The lack of significant evidence for curvature in the VHE spectrum, coupled with the observed harder when brighter behaviour some HBLs have displayed, makes future flaring activity for this source of great interest due the hard deabsorbed spectral index. 
The spectral index obtained indicates that the higher-energy peak of the broadband SED (i.e. the inverse-Compton component) is near or below the lowest-energy VERITAS point at 200 GeV.
 
 
While only marginal evidence is seen in the VERITAS data, curvature or a cutoff is expected to occur at some point in the intrinsic VHE spectrum of a BL Lac object if the emission is produced by the inverse-Compton mechanism. In one-zone SSC models, a single population of relativistic electrons with a spectrum that softens with increasing energy is utilised to produce a synchrotron spectrum that matches the observed optical-X-ray SED. Commonly this electron distribution is described by a broken power-law 
between Lorentz factors $\gamma_\text{min}$ and $\gamma_\text{max}$ with a break at $\gamma_\text{b}$ \citep[e.g.][]{katarztnski2001ssc}. This is in turn Doppler boosted by a factor $\delta/(1+z)$, where $\delta = \gamma^{-1} (1 - \beta\cos\theta)$ and $\beta$ is the relativistic velocity, to higher frequencies in the frame of the observer by the bulk relativistic motion of the jet. The resulting synchrotron spectrum follows power laws for corresponding frequencies between $\gamma_\text{min}$ and $\gamma_\text{b}$ and between $\gamma_\text{b}$  and $\gamma_\text{max}$, with the latter power-law index being softer to account for cooling and escape of the electrons. 
The frequency range of the synchrotron radiation produced depends on the strength of the magnetic field and $\gamma$ (the synchrotron critical frequency for a single electron $\nu_{ch}\propto \gamma^2 B$), but decreases sharply beyond that corresponding to $\gamma_\text{max}$, hence terminating the synchrotron spectrum.  Typically the electrons boost, via the inverse-Compton mechanism, the energy of the synchrotron photons they produced by a factor of $\gamma^2$, with the interaction rate governed approximately by the Thomson cross section. For combinations of large electron Lorentz factors and high target synchrotron photon frequencies, the energies of the synchrotron photons in the electron rest frame may be comparable to the electron rest mass. This limits the boost in energy to $\gamma m_0 c^2$ and the Klein-Nishina cross section applies, which results in a decreased interaction rate and a softening of the inverse-Compton spectrum at the highest energies. Thus, there is a variety of combinations of electron and soft photon energies that could produce the VHE emission and result in a particular spectral shape over the range in which we have measured it. It is difficult to constrain the models from observations in one band alone, and constraints on the model parameter space require contemporaneous measurements of both the synchrotron and inverse-Compton components and any associated time lags between variability in the different emission bands. While many models involve numerical calculations, analytic approaches to constraining parameters can be found in, for example, \citet{tavecchio1998constraints} .

VERITAS will continue to monitor 1ES~2344+514 as part of the blazar science working group's long-term plan over the coming years and will provide and respond to alerts should flaring activity occur. In the absence of flaring activity, continued observations will increase the photon statistics available to extend the spectrum to higher energies to allow for models of the emission of BL Lac objects in quiescent states to be constrained, and will allow further studies of the characteristics and implications of its temporal, spatial and spectral properties to be conducted.

Being one of the first VHE-detected blazars, 1ES~2344+514 receives significant multiwavelength monitoring from instruments across the electromagnetic spectrum. The VERITAS data presented here represent the most-comprehensive VHE study of this object to date. The VERITAS observations have significant contemporaneous optical (e.g. Tuorla\footnote{\url{http://users.utu.fi/kani/1m/}}) and Swift XRT X-ray ($>$144 exposures since September 2007\footnote{\url{http://www.swift.psu.edu/monitoring/data/1ES2344+514/lightcurve.txt}}, many of which taken simultaneously with VERITAS) data available, in addition to the almost-continuous {\it Fermi}-LAT coverage. The combining of these data sets, along with others from additional instruments, can prove useful for understanding the underlying processes in 1ES~2344+514, and HBLs in general, and is encouraged.

\section*{Acknowledgements}
This research is supported by grants from the U.S. Department of Energy Office of Science, the U.S. National Science Foundation and the Smithsonian Institution, and by NSERC in Canada. We acknowledge the excellent work of the technical support staff at the Fred Lawrence Whipple Observatory and at the collaborating institutions in the construction and operation of the instrument. S. O'Brien is funded by a UCD Research Scholarship. The VERITAS Collaboration is grateful to Trevor Weekes for his seminal contributions and leadership in the field of VHE gamma-ray astrophysics, which made this study possible.





\bibliographystyle{mnras} 
\bibliography{References} 






\bsp	
\label{lastpage}
\end{document}